\documentstyle[12pt]{article}
\newcommand{\beq}{\begin{equation}}
\newcommand{\eeq}{\end{equation}}
\newcommand{\beqa}{\begin{eqnarray}}
\newcommand{\eeqa}{\end{eqnarray}}
\newcommand{\ba}{\begin{array}}
\newcommand{\ea}{\end{array}}
\newcommand{\CR}{\nonumber \\}

\newcommand{\half}{{1\over 2}}
\newcommand{\Tr}{{\rm Tr}}
\newcommand{\tr}{{\rm tr}\,}

\newcommand{\wt}{\widetilde}
\newcommand{\wh}{\widehat}

\newcommand{\bA}{{\bf A}}
\newcommand{\bB}{{\bf B}}
\newcommand{\bC}{{\bf C}}
\newcommand{\bD}{{\bf D}}
\newcommand{\bE}{{\bf E}}
\newcommand{\bH}{{\bf H}}
\newcommand{\bP}{{\bf P}}
\newcommand{\bR}{{\bf R}}
\newcommand{\bX}{{\bf X}}
\newcommand{\bZ}{{\bf Z}}
\newcommand{\cN}{{\cal N}}
\def\br#1#2{\bX_{\bf [#1,#2]}}
\def\X#1#2{\bX_{\bf [#1,#2]}}
\def\K#1#2{K_{[#1,#2]}}
\def\E#1{\bE_{\bf #1}}
\def\htE#1{\widehat{\widetilde \bE}_{\bf #1}}

\begin{document}

\begin{titlepage}
\begin{flushright}
{\tt hep-th/9907134} \\
UTHEP-407 \\
July, 1999
\end{flushright}
\vspace{0.5cm}
\begin{center}
{\Large \bf 
Affine 7-brane Backgrounds and \\
Five-Dimensional $E_N$ Theories on $S^1$
\par}
\lineskip .75em
\vskip2.5cm
\normalsize
{\large Yasuhiko Yamada}
\vskip 1.5em
{\large\it Department of Mathematics, Kobe University \\
Rokko, Kobe 657-8501, Japan}
\vskip1cm
{\large Sung-Kil Yang}
\vskip 1.5em
{\large\it Institute of Physics, University of Tsukuba \\
Ibaraki 305-8571, Japan}
\end{center}
\vskip3cm
\begin{abstract}
Elliptic curves for the 7-brane configurations realizing the affine Lie 
algebras $\wh E_n$ $(1 \leq n\leq 8)$ and $\wh{\wt E}_n$ $(n=0,1)$ are 
systematically derived from the cubic equation for a rational elliptic 
surface. It is then shown that the $\wh E_n$ 7-branes describe the 
discriminant locus of the elliptic curves for five-dimensional (5D) $\cN =1$ 
$E_n$ theories compactified on a circle. This is in accordance with a recent 
construction of 5D $\cN =1$ $E_n$ theories on the IIB 5-brane web with 
7-branes, and indicates the validity of the D3 probe picture for 5D 
$E_n$ theories on $\bR^4 \times S^1$. Using the $\wh E_n$ curves we also study
the compactification of 5D $E_n$ theories to four dimensions.
\end{abstract}

\end{titlepage}

\baselineskip=0.7cm

\section{Introduction}

In a series of papers \cite{Joh}-\cite{DHIZ-3} the 7-brane technology has been 
developed systematically in connection with the F-theory compactification on
an elliptic $K3$ surface. It is particularly interesting that there exist
the 7-brane configurations on which infinite symmetries of the affine Lie
algebras $\wh E_N$ are realized \cite{dW,DHIZ-3}.

An interesting application of the 7-brane technology has been found recently 
by Hanany et al. in constructing five-dimensional (5D) $\cN =1$ theories
on the brane web \cite{DHIK}. 
In 5D, Seiberg discovered non-trivial interacting $\cN =1$
superconformal theories with global $E_N$ symmetries \cite{Sei}.
It is quite difficult to obtain such $E_N$ theories in a conventional
M-theory description of the brane web. In \cite{DHIK}, however, 
it is shown that introducing 7-branes in the $(p,q)$ 5-brane web makes it 
possible to construct 5D $\cN =1$ theories with $E_N$ 
symmetries on the web. In this approach 
the affine property of the 7-branes plays a crucial role. 

5D $\cN =1$ $E_N$ theories are known to arise upon compactifying M-theory on a 
Calabi-Yau threefold with a shrinking del Pezzo four-cycle \cite{MS,DKV}.
Minahan et al. obtained the elliptic curves for the
Coulomb branch of 5D $E_N$ theories compactified on 
a circle \cite{MNW}. Their construction of the curves is motivated by
considering the local model of a singular Calabi-Yau threefold where a 
del Pezzo surface shrinks to zero size \cite{LMW}. 

Since the 7-brane configurations are described in terms of elliptic 
curves \cite{Vafa} we expect that the curves for 5D $E_N$ theories may be
derived from the 7-branes if their affine property is properly taken into
account. We will find that this is indeed the case.

In this paper, after briefly reviewing the 7-branes with exceptional symmetries
in section 2, we construct elliptic curves to describe the 
7-branes realizing the affine algebras $\wh E_N$ $(1 \leq N \leq 8)$ and 
$\wh{\wt E}_N$ $(N=0,1)$ in section 3.
We then show in section 4 that our curves coincide with those obtained 
in \cite{MNW} for 5D $E_N$ theories. We also discuss the compactification 
of 5D $E_N$ theories down to four dimensions by reducing the curves for 5D
theories to the known Seiberg-Witten curves for four-dimensional (4D) $\cN =2$
theories.

\section{Affine 7-branes}

\renewcommand{\theequation}{2.\arabic{equation}}\setcounter{equation}{0}

The configurations of IIB 7-branes are described in terms of an elliptic
curve
\beq
y^2=x^3+f(z) x+g(z),
\label{cubic}
\eeq
where $f$ and $g$ are certain polynomials in $z$, 
and $z$ is a complex coordinate of ${\bf P}^1$ which is the space 
transverse to the 7-branes  \cite{Vafa}. 
The zeroes of the discriminant
\beq
\Delta (z)=4f(z)^3+27g(z)^2
\label{discri}
\eeq
determine the transverse positions of the 7-branes. 
The $SL(2,\bZ)$ invariant is defined by
\beq
J={4 f(z)^3 \over \Delta(z)}.
\eeq
When $f$ is of degree $\leq 4k$ and $g$ is of degree $\leq 6k$,
the cubic (\ref{cubic}) defines a rational elliptic surface for $k=1$ and
an elliptic $K3$ surface for $k=2$. 
Let ${\rm I}_{p,q}$ denote a type ${\rm I}_1$ fiber specified by a
vanishing cycle $p\alpha +q\beta$ where $p,q$ are mutually prime integers
and $\alpha ,\beta$
are homology cycles of a fiber torus. 
A fiber of type ${\rm I}_1$ has the $A_0$ singularity and its
monodromy is given by $SL(2,\bZ)$ conjugate to 
\beq
T=\pmatrix{1&1 \cr 0&1\cr} .
\eeq
Thus there exists generically a fiber ${\rm I}_{p,q}$ 
at the transverse position 
of a 7-brane. Following \cite{DHIZ-2,DHIZ-3} we will label such a 7-brane 
as $\br{p}{q}$.

Type IIB $(p,q)$ strings can end on a 7-brane $\br{p}{q}$. 
Since $(p,q)$ strings 
are obtained from the fundamental $(1,0)$ strings by the $SL(2,\bZ)$ 
transformation the monodromy matrix $\K{p}{q}$ associated to $\br{p}{q}$
turns out to be\footnote[2]{Following the convention of \cite{DHIZ-2,DHIZ-3}
the monodromy matrix $K_{[p,q]}$ is the inverse of the usual monodromy matrix.}
\beq
\K{p}{q} =\pmatrix{1+pq & -p^2 \cr q^2 & 1-pq \cr}.
\eeq
Given a 7-brane configuration 
$\bX_{\bf 1} \cdots \bX_{\bf n-1}\bX_{\bf n}$ we have the
total monodromy matrix $K=K_nK_{n-1} \cdots K_1$. The value of $\tr K$ is
relevant to the classification of 7-brane configurations since $\tr K$ is
an $SL(2,\bZ)$ conjugation invariant \cite{DHIZ-2,DHIZ-3}.
Classification of collapsible brane configurations is nothing but the 
celebrated Kodaira classification of singular fibers \cite{Kod}. 
Denoting $\br{1}{0}=\bA, \br{1}{-1}=\bB, \br{1}{1}=\bC$
we list the correspondence between the Kodaira and brane
classifications in Table 1. All the brane configurations in Table 1 have the
monodromy $K$ with $|\tr K| \leq 2$.

\renewcommand{\arraystretch}{1.2}
\begin{table}
\begin{center}
\begin{tabular}{||c|c|c|c||} \hline
fiber    & singularity  & 7-branes & brane \\
type     & type         &          & type  \\ \hline\hline
${\rm I}_n$    &  $A_{n-1}$   & $\bA^n$     &  $\bA_{\bf n-1}$   \\
${\rm II}$     &  $A_0$       & $\bA\bC$    &  $\bH_{\bf 0}$   \\
${\rm III}$    &  $A_1$       & $\bA^2\bC$  &  $\bH_{\bf 1}$   \\
${\rm IV}$     &  $A_2$       & $\bA^3\bC$  &  $\bH_{\bf 2}$   \\
${\rm I}_0^*$  &  $D_4$       & $\bA^4\bB\bC$   &  $\bD_{\bf 4}$   \\
${\rm I}_n^*$  &  $D_{n+4}$   & $\bA^{n+4}\bB\bC$   &  $\bD_{\bf n+4}$   \\
${\rm II}^*$   &  $E_8$       & $\bA^7\bB\bC^2$     &  $\bE_{\bf 8}$   \\
${\rm III}^*$  &  $E_7$       & $\bA^6\bB\bC^2$     &  $\bE_{\bf 7}$   \\
${\rm IV}^*$   &  $E_6$       & $\bA^5\bB\bC^2$   &  $\bE_{\bf 6}$   \\  \hline
\end{tabular}
\end{center}
\caption{Kodaira classification, ADE singularities and 7-branes. 
$n \geq 1$ for ${\rm I}_n$ and ${\rm I}_n^*$}
\label{tbl1}
\end{table}

The collapsible 7-branes are of physical importance since they represent 
the enhanced gauge symmetries in the bulk. The string/junction configurations
giving rise to the massless gauge bosons are found in \cite{dWZ}. 
This is done by 
identifying the BPS junctions with the root vectors of the finite Lie algebra
which is determined by the singularity type in Table 1. It is also 
important to consider non-collapsible 7-brane configurations beyond
the Kodaira classification. For instance $\bD_{\bf n}=\bA^n\bB\bC$ with
$0\leq n \leq 3$ are relevant when we describe the Seiberg-Witten solution
for four-dimensional $N=2$ $SU(2)$ QCD with $N_f=n$ fundamental flavors
in the D3 probe picture \cite{Sen,BDS}. 

Extending the list in Table 1 to non-collapsible configurations 
there appear infinite series of 7-brane
configurations \cite{DHIZ-2} among which we are particularly interested 
in the $E_N$ and $\widetilde E_N$ series defined by
\beq
\E{N}=\bA^{N-1}\bB\bC^2, \hskip10mm \widetilde \E{N}=\bA^N\br{2}{-1}\bC.
\label{Ebrane}
\eeq
It should be noted, however, that these two series are equivalent for 
$N \geq 2$, while $\E{1}=\bA\bB\bC^2$ and $\widetilde \E{1}=\bA\br{2}{-1}\bC$
are not equivalent \cite{DHIZ-2}. 
The finite Lie algebras associated to $\E{N}$ are
easily identified once we see how $\E{8,7,6}$ yield the root vectors of
$E_{8,7,6}$. Corresponding to $\E{N}$ with $N \leq 5$ we have $E_5=D_5$,
$E_4=A_4$, $E_3=A_1 \oplus A_2$, $E_2=A_1\oplus u(1)$ and $E_1=A_1$. 
Likewise one finds $\widetilde E_1=u(1)$ for $\widetilde \E{1}$ and
no symmetry for $\widetilde\E{0}$. $\E{N}$ with $N \geq 9$ do not realize 
the finite Lie algebra.

The series (\ref{Ebrane}) are very interesting since they admit the affine
extension \cite{dW,DHIZ-3}. The affine configurations are constructed 
by adding a single 7-brane to (\ref{Ebrane})
\beq
\widehat\E{N}=\bA^{N-1}\bB\bC^2\X{3}{1}, \hskip10mm 
\htE{N}=\bA^N\br{2}{-1}\bC\X{4}{1},
\label{Ehat}
\eeq
where $\widehat\E{N}$ and $\htE{N}$ are equivalent for $N\geq 2$ as in the
finite case. Note also the equivalence relation
\beq
\widehat\E{N}=\bA^{N-1}\bB\bC^2\X{3}{1} =\bA^{N-1}\bB\bC\bB\bC .
\label{canoE8}
\eeq
There now exists a BPS loop junction which goes around the
7-branes. This junction is identified with the imaginary root so that the
root system associated to (\ref{Ehat}) turns out to be the affine root system.
Thus the 7-branes (\ref{Ehat}) realize infinite dimensional symmetries
of the affine algebras $\widehat E_N$ and $\widehat{\widetilde E}_N$.
Note that the monodromy matrices for $\widehat\E{N}$ and $\htE{N}$ are
identical
\beq
K(\widehat\E{N})=K(\htE{N})=\pmatrix{1 & 9-N \cr 0 & 1 \cr}.
\eeq
We refer to \cite{DHIZ-2,DHIZ-3} for further extensive explanation of
7-brane configurations.

\section{Elliptic curves for affine 7-branes}

\renewcommand{\theequation}{3.\arabic{equation}}\setcounter{equation}{0}

\renewcommand{\arraystretch}{1.7}
\begin{table}
\begin{center}
\begin{tabular}{||c|c|c||} \hline
$\htE{0}$   &   $\wh\E{1}$  &  $\htE{1}$, $\wh\E{N}$ $(2 \leq N\leq 8)$ \\
\hline
$\bP^2$   &   $\bP^1 \times \bP^1$  &  $\bP^2$ with the $N$ points blown up \\
\hline
\end{tabular}
\end{center}
\caption{7-branes and del Pezzo surfaces}
\label{tbl2}
\end{table}
\renewcommand{\arraystretch}{1}
Let us concentrate on $\wh\E{N}$ with $N \leq 9$, $\htE{1}$ 
and $\htE{0}$. These exceptional series of 7-branes have an intimate relation
with the del Pezzo surface \cite{DHIK}. 
The correspondence is summarized in Table 2.
It is not difficult to figure out how a basis for the two-cycles in the del 
Pezzo is translated into a basis of the junction lattice on the corresponding
7-branes. On top of this, the $\wh\E{9}$ configuration, which contains
12 7-branes, corresponds to the ninth del Pezzo surface which is an elliptic 
manifold over $\bP^1$ described by (\ref{cubic}) with
\beq
f(z)=\sum_{i=0}^4 a_iz^i, \hskip10mm g(z)=\sum_{i=0}^6 b_iz^i.
\label{ratpol}
\eeq
The ninth del Pezzo, dubbed also $\half K3$, has appeared in the
literature on the F-theory description of six-dimensional non-critical
strings \cite{MV}-\cite{MNVW}. This surface is not the blow-up of $\bP^2$ 
at generic nine points, but the position of the ninth point is fixed by 
the other eight points.

Starting with the $\wh E_9$ curve (\ref{cubic}) with (\ref{ratpol})
we now wish to construct $\wh E_N$
curves for $N\leq 8$ in a systematic way. For generic values of $a_i, b_i$
the 7-branes are located in the finite region on the $z$-plane. Our basic
idea to obtain $\wh\E{N}$ is to remove $(9-N)$ $\bA$-branes from $\wh\E{9}$ 
and send them infinitely far away from the rest. If we place the removed 
7-branes $\bA^{9-N}$ at $z=0$, they are described as the $A_{8-N}$ singularity
at $z=0$ (see Table \ref{tbl1}). The appearance of the series of $A$-type
singularities makes the derivation of $\wh E_N$ curves systematic.
In fact, according to the Tate algorithm, the $\wh E_9$
curve should obey the condition
\beq
{\rm ord}_z(J) =-(9-N), \hskip10mm {\rm ord}_z(f)\equiv 0 \quad 
({\rm mod}\; 2),
\eeq
where ${\rm ord}_z(\bullet)$ means the vanishing order of $\bullet$ in $z$.
Writing the discriminant as
\beq
\Delta =\sum_{i=0}^{12} \Delta_i(a,b)z^i,
\eeq
we see that the condition is satisfied by imposing $\Delta_i=0$ for 
$0 \leq i \leq 8-N$. Then we set $z=1/u$ and $(x,y) \rightarrow (u^4x, u^6y)$ 
to express the curve in such a way that the $\wh\E{N}$ 
branes are located around $u=0$.

First of all, to obtain the $\wh E_8$ curve requires
\beq
\Delta_0 =4a_0^3+27 b_0^2 =0
\eeq
{}from which we take
\beq
a_0=-3/L^4, \hskip10mm b_0=2/L^6.
\label{b0}
\eeq
Here $L$ is regarded as a scale parameter which will play an important role 
later. Thus the generic $\wh E_8$ curve reads
\beq
y^2=x^3+\left( -{3u^4 \over L^4}+\sum_{i=0}^3a_{4-i}u^i \right) x
+\left( {2u^6\over L^6}+\sum_{i=0}^5 b_{6-i}u^i \right).
\label{massE8}
\eeq
Next, the $\wh E_7$ curve is obtained from $\Delta_0=0$ as well as 
$\Delta_1=0$. The latter yields
\beq
b_1=-a_1/L^2.
\label{b1}
\eeq
The generic $\wh E_7$ curve is then written as
\beq
y^2=x^3+\left( -{3u^4 \over L^4}+\sum_{i=0}^3a_{4-i}u^i \right) x
+\left( {2u^6\over L^6}-{a_1 \over L^2}u^5+\sum_{i=0}^4 b_{6-i}u^i \right).
\eeq

Repeating this procedure we see that $\Delta_j=0$ for $2\leq j \leq 6$
vanish if and only if
\beqa
&& b_2=\frac{1}{12 L^2}\left(a_1^2 L^4 -12 a_2 \right), \CR 
&& b_3=\frac{1}{216 L^2}\left(a_1^3 L^8+36 a_1 a_2 L^4 -216 a_3 \right), \CR 
&& b_4=\frac{1}{1728 L^2}
\left(a_1^4 L^{12}+24 a_1^2 a_2 L^8
+144 a_2^2 L^4+288 a_1 a_3 L^4-1728 a_4 \right), \CR
&& b_5=\frac{L^2}{10368}
\left(a_1^5 L^{12}+24 a_1^3 a_2 L^8+144 a_1 a_2^2 L^4+144 a_1^2 a_3 L^4+
1728 a_2 a_3+1728 a_1 a_4 \right),  \CR  
&& b_6=\frac{L^2}{373248}
\left(7 a_1^6 L^{16}+180 a_1^4 a_2L^{12}+1296 a_1^2 a_2^2 L^8+
864 a_1^3 a_3 L^8+1728 a_2^3L^4  \right.  \CR 
&&   \hskip25mm   \left. +10368 a_1 a_2 a_3 L^4+5184 a_1^2 a_4 L^4+
31104 a_3^2+62208 a_2 a_4 \right).
\label{bi}
\eeqa
Thus the generic curves of type $\wh E_N$ for $N=6,5,\cdots ,2$ are explicitly
written down by substituting (\ref{b1}), (\ref{bi}) for $b_i$
$(0\leq i \leq 8-N)$ into (\ref{massE8}).

An amusing phenomenon occurs when we require $\Delta_7=0$. We find $\Delta_7$
in the factorized form, and hence $\Delta_7=0$ yields either
\beq
a_3=-\frac{L^4}{72}a_1 \left(a_1^2L^4 +12 a_2 \right)
\label{a3s}
\eeq
or
\beq
a_4=-\frac{L^4}{576}
\left(a_1^4 L^8+16 a_1^2 a_2 L^4+48 a_2^2 +48 a_1 a_3 \right).
\label{a4s}
\eeq
Two curves obtained from (\ref{a3s}) and (\ref{a4s}) are identified with
the generic $\wh E_1$ and $\wh{\wt E}_1$ curves, respectively. Branching of
the sequence at rank one nicely corresponds to the properties of 
$\wh\E{N}$ and $\htE{N}$ branes at $N=1$. 

Inspecting the discriminant, it is observed that the $\wh{\wt E}_1$ curve
admits the further degeneration, {\it i.e.} the $A_8$ singularity at 
$u=\infty$. We find that putting
\beq
a_2=-\frac{L^4}{8} a_1^2
\eeq
in the $\wh{\wt E}_1$ curve gives the $\wh{\wt E}_0$ curve.

We note here that the curves obtained so far admit the $GL(2, \bC)$
transformation
\beq
u \rightarrow \frac{c_1 u+c_2}{c_3 u+c_4}, \hskip10mm
x \rightarrow \frac{x}{(c_3 u+c_4)^2}, \hskip10mm
y \rightarrow \frac{y}{(c_3 u+c_4)^3}.
\label{sl2tr}
\eeq
By virtue of this, let us see that the expressions of the curves for
$\wh E_1, \wh {\wt E}_1, \wh {\wt E}_0$ are greatly simplified. 
In the $\wh E_1$ curve, in view of the transformation (\ref{sl2tr}) we put
\beqa
&& a_1=\frac{12}{L^3 p} (16+p^2), \hskip10mm
a_2=-\frac{6}{L^2 p^2} (512-160 p^2+3 p^4),  \CR
&& a_4=-\frac{3}{p^2}(-192+p^2)(1024-256 p^2+p^4).
\eeqa
We next apply (\ref{sl2tr}) with 
$c=(c_1,c_2,c_3,c_4)=(\frac{L}{\sqrt{3}},0,0,\frac{1}{\sqrt{3}})$ and 
shift $x$. The $\wh E_1$ curve is then expressed as
\beq
y^2=x^3+\left(u^2-2 u \left(p+\frac{16}{p}\right)+p^2-224 \right)x^2
+\frac{65536}{p^2} x.
\label{massE1at}
\eeq

In the $\wh{\wt E}_1$ curve, we put
\beqa
&& a_1=\frac{12}{L^3 p} (16+p^2), \hskip10mm
a_2=-\frac{6}{L^2 p^2} (512+96 p^2+3 p^4),\CR
&& a_3=\frac{12}{L p} (3584+48 p^2+p^4).
\eeqa
We next apply (\ref{sl2tr}) with
$c=(\frac{L}{\sqrt{3}},0,0,\frac{1}{\sqrt{3}})$ and shift $x$.
The $\wh{\wt E}_1$ curve is then expressed as
\beq
y^2=x^3+\left(u^2-2 u \left(p+\frac{16}{p} \right)+32+p^2 \right)x^2+
4096 \left(\frac{u}{p}-\frac{16}{p^2}-1 \right)x+\frac{4194304}{p^2}.
\label{masstE1at}
\eeq

In the $\wh{\wt E}_0$ curve, we put
\beq
a_1=0, \hskip10mm
a_3=-\frac{8}{3L}.
\eeq
We next apply (\ref{sl2tr}) with
$c=(\sqrt{3} L,0,0,\sqrt{3})$ and shift $x$.
The curve now reads
\beq
y^2=x^3+9 u^2 x^2-24 u x+16
\eeq
with the discriminant $\Delta =6912 (u^3+1)$. Three zeroes of $\Delta$ at
$u=e^{i\pi (2j+1)/3}$ with $j=0,1,2$ are the positions of three 7-branes
of $\htE{0}$. Evaluating the monodromy $M_j$ at $u=e^{i\pi (2j+1)/3}$ we
obtain $M_0=K_{[1,1]}, M_1=K_{[-1,2]}, M_2=K_{[2,-1]}$. This is in agreement
with the monodromy of $\htE{0}=\X{2}{-1}\bC\X{4}{1}=\X{2}{-1}\X{-1}{2}\bC$.
The $\bZ_3$ symmetry acting on the $u$-plane is observed by noting that
$STM_j(ST)^{-1}=M_{j+1\, ({\rm mod}\; 3)}$ where 
$S=\pmatrix{0 & -1 \cr 1 & 0 \cr}$ and $(ST)^3=1$.

To summarize, our derivation of the $\wh E_N$ as well as $\wh{\wt E}_N$
curves is depicted in the diagram
\def\rr{\rightarrow}
\def\ne{\nearrow}
\def\se{\searrow}
\beq
\begin{array}{ccccccccccccccccccccccc}
&&&&&&&&&&&&&&&&A_7&\rr&A_8 \\
&&&&&&&&&&&&&&&\ne \\
0&\rr&A_0&\rr&A_1&\rr&A_2&\rr&A_3&\rr&A_4&\rr&A_5&\rr&A_6&\rr&A_7 \\
&&&&&&&&&&&&&&&&& \\
&&&&&&&&&&&&&&&&\wh{\wt{E}}_1&\rr&\wh{\wt{E}}_0 \\
&&&&&&&&&&&&&&&\ne \\
\wh{E}_9&\rr&\wh{E}_8&\rr&\wh{E}_7&\rr&\wh{E}_6&\rr&\wh{E}_5&
\rr&\wh{E}_4&\rr&\wh{E}_3&\rr&\wh{E}_2&\rr&\wh{E}_1 
\end{array}
\eeq
Here the upper sequence stands for the coalescence flow of the 
$A$-type singularities ({\it i.e.} $\bA^{8-N}$-branes) at $u=\infty$ 
while the lower sequence stands for the degeneration flow of the 
$\wh \E{N}$ configurations.
As we will see in section 4 this diagram corresponds to the renormalization 
group flows among 5D $\cN =1$ supersymmetric $E_N$ theories.

\subsection{Massless curves}

In order to see more explicitly the correspondence between the $\wh E_N$
curves and the $\wh \E{N}$ branes we now study the degeneration of
the curves. For each $\wh \E{N}$ there is a maximally collapsible
sub-configuration of 7-branes \cite{DHIK}.
Correspondingly the $\wh E_N$ curve has a degeneration which
realizes the $E_N$ singularity.
We describe such degeneration by adjusting the parameters $a_i, b_i$. 
The curves obtained in this way will be referred to as the massless curves. 
Our results are then as follows:

\vskip3mm\noindent
$\bullet$ Massless $\wh E_8$:

In order to have the $E_8$ singularity at $u=0$,
we must set $a_i=0$ ($i=4,3,2,1$) and $b_i=0$ ($i=6,5,4,3,2$).
Then we obtain the massless $\wh E_8$ curve
\beq
y^2=x^3-{3 \over L^4}u^4 x+\left({2\over L^6}u^6 +b_1 u^5 \right)
\eeq
with the discriminant
\beq
\Delta =27b_1u^{10}\left( 4u+b_1L^6 \right)/L^6.
\eeq
The form of $\Delta$ indicates that 10 zeroes at $u=0$ represent the coalescing
$\E{8}=\bA^7\bB\bC^2$ branes and a zero at $u=-b_1L^6/4$ a 7-brane $\X{3}{1}$.
Since $\wh\E{8}$ is not collapsible $\X{3}{1}$ keeps a finite distance from
the collapsible $\E{8}$.

\vskip3mm\noindent
$\bullet$ Massless $\wh E_7$:

Setting $a_i=0$ ($i=4,3,2$) and $b_i=0$ ($i=6,5,4,3,2$) we obtain
\beq
y^2=x^3+\left(-{3 \over L^4}u^4 +a_1u^3\right) x
+\left({2\over L^6}u^6 -{a_1\over L^2} u^5 \right)
\eeq
with the discriminant
\beq
\Delta =-a_1^2u^9 \left( 9u-4a_1L^4 \right)/L^4.
\eeq
Thus there exist coalescing $\E{7}=\bA^6\bB\bC^2$ branes at $u=0$ 
and $\X{3}{1}$ at $u=4a_1L^4/9$.

\vskip3mm\noindent
$\bullet$ Massless $\wh E_6$:

Setting $a_i=0$ ($i=4,3,2$) and $b_i=0$ ($i=6,5,4,3$) we obtain
\beq
y^2=x^3+\left(-{3 \over L^4}u^4 +a_1u^3\right) x
+\left({2\over L^6}u^6 -{a_1\over L^2} u^5+{a_1^2L^2 \over 12}u^4 \right)
\eeq
with the discriminant
\beq
\Delta =-a_1^3u^8 \left( 8u-3a_1L^4 \right)/16.
\eeq
Thus there exist coalescing $\E{6}=\bA^5\bB\bC^2$ branes at $u=0$ 
and $\X{3}{1}$ at $u=3a_1L^4/8$.

\vskip3mm\noindent
$\bullet$ Massless $\wh E_5$:

Setting $a_i=0$ ($i=4,3$), $a_2=-\frac{L^4}{48}a_1^2$ and $b_i=0$ ($i=6,5,4$) 
we obtain
\beq
y^2=x^3+\left(-{3 \over L^4}u^4 +a_1u^3-{L^4a_1^2 \over 48}u^2 \right) x
+\left({2\over L^6}u^6 -{a_1\over L^2} u^5+{5 \over 48}a_1^2L^2 u^4
+{a_1^3L^6 \over 864}u^3 \right)
\eeq
with the discriminant
\beq
\Delta =-3 a_1^4 L^4 u^7 \left( 3u-a_1L^4 \right)/256.
\eeq
Thus there exist 7 coalescing 7-branes at $u=0$ which realize $E_5=D_5$. 
This is seen from the equivalence 
\beq
\wh \E{5}=\bA^4\bB\bC^2\X{3}{1}=\bA^5\X{2}{-1}\bC\X{4}{1}
=\bD_{\bf 5}\X{4}{1},
\eeq
where $\X{4}{1}$ is located at $u=a_1L^4/3$.

\vskip3mm\noindent
$\bullet$ Massless $\wh E_4$:

Here we encounter a somewhat different situation from the above.
Writing the discriminant of the $\wh E_4$ curve as
\beq
\Delta =\sum_{i=0}^7 \Delta_i(a,b)u^i
\eeq
we first require $\Delta_0=4a_4^3+27b_6^2=0$. This is obeyed by taking
\beq
a_4=-3/T^4, \hskip10mm b_6=2/T^6,
\eeq
where, in addition to $L$, another scale parameter $T$ has appeared. We can 
then set at most $\Delta_j=0$ for $0 \leq j \leq 4$ by tuning
\beqa
&& b_5=-{a_3 \over T^2},\hskip10mm a_1={36 \over L^3T}, \hskip10mm
a_3=-{2 \over LT^3}\left(a_1L^3T-18\right), \CR
&& a_2=-{1\over 12 L^2T^2}\left( a_1^2L^6T^2+12a_3LT^3-12a_1L^3T+72 \right),
\eeqa
{}from which the massless curve is obtained as 
\beqa
&& y^2=x^3+\left(-{3 \over L^4}u^4 +{36 \over L^3T}u^3-{42 \over L^2T^2}u^2
-{36 \over LT^3}u-{3\over T^4} \right) x   \CR
&& \hskip10mm +\left({2\over L^6}u^6 -{36\over L^5T} u^5+{150 \over L^4T^2} u^4
+{150 \over L^2T^4}u^2+{36 \over LT^5}u+{2 \over T^6} \right)
\eeqa
with the discriminant
\beq
\Delta =-186624 u^5 \left( T^2u^2-11LTu-L^2 \right)/(LT)^7.
\eeq
The existence of 5 coalescing 7-branes at $u=0$ is understood by noting that
$\wh \E{4}=\bA^3\bB\bC^2\X{3}{1}$ is equivalent to $\bB^5\X{-2}{3}\bC$.
Hence we have the $A_4$ singularity at the origin in agreement with the
relation $E_4=A_4$. Two 7-branes $\X{-2}{3}$ and $\bC$ stay at a 
distance from the collapsed $\bB^5$ at $u=0$, reflecting the fact that 
both $\wh \E{4}$ and $\E{4}$ are non-collapsible configurations.

\vskip3mm\noindent
$\bullet$ Massless $\wh E_3$:

We proceed in parallel with the $\wh E_4$ case. The maximal degeneracy of the
curve is achieved by taking
\beq
 a_4=-{3 \over T^4},\hskip10mm b_6={2 \over T^6},\hskip10mm
a_3=-{6 \over LT^3}, \hskip10mm a_2=0, \hskip10mm a_1={12 \over L^3T}.
\eeq
The massless curve turns out to be
\beqa
&& y^2=x^3+\left(-{3 \over L^4}u^4 +{12 \over L^3T}u^3
-{6 \over LT^3}u-{3\over T^4} \right) x   \CR
&& \hskip10mm +\left({2\over L^6}u^6 -{12 \over L^5T} u^5+{12 \over L^4T^2} u^4
+{14 \over L^3T^3}u^3+{3 \over L^2T^4}u^2+{6 \over LT^5}u+{2 \over T^6} \right)
\eeqa
with the discriminant
\beq
\Delta =-729 u^3 \left( uT-4L \right) \left(L+2uT \right)^2/(L^6T^9).
\eeq
As in the previous case, the structure of the discriminant is again 
understood by showing that
$\wh \E{3}=\bA^2\bB\bC^2\X{3}{1}$ is equivalent to 
$\X{2}{-1}^2\X{-1}{1}^3\bC$ in accordance with the fact that 
$E_3=A_1\oplus A_2$. 

\vskip3mm\noindent
$\bullet$ Massless $\wh E_2$:

We need to have the $A_1$ singularity at $u=0$. For this the coefficients
of $u$ and $u^0$ in the discriminant should vanish. At first sight it looks
quite difficult to find such a solution. If we eliminate $a_4$ from these two
equations, however, the resulting expression takes a remarkable factorized
form. As a result we see that $a_3$ and $a_4$ should be 
\beqa
&& a_3=\frac{L^4}{576} a_1 \left(a_1^2 L^4 - 24 a_2 \right), \CR
&& a_4=-\frac{L^4}{27648} 
\left(25 a_1^4 L^8+ 240 a_1^2 a_2L^4 + 576 a_2^2 \right).
\eeqa
Rescaling the variables as $a_2=\frac{k}{24} a_1^2 L^4$,
$u \rightarrow \frac{1}{24} a_1 L^4$ and
$x \rightarrow \frac{1}{576} a_1^2 L^6$,
we find the massless curve in the form
\beqa
y^2&=&x^3+(-12 (k+5)^2-24 (k-1)u+24 k u^2+24 u^3-3 u^4)x \CR
&& +16 (k+5)^3+48(k-1)(k+5)u+12(49+18k+5 k^2)u^2 \CR
&& +40(3k+1)u^3-24(k-2)u^4-24 u^5+2 u^6
\label{E2withk}
\eeqa
with the discriminant
\beq
\Delta =11664(k+3)^3 u^2(200+80k+8 k^2+(72+40k)u+(21-k)u^2-4 u^3).
\eeq
The symmetry on the brane is expected to be $E_2=A_1 \oplus u(1)$. In fact,
two zeroes at $u=0$ represent two collapsed $\bC$-branes in
$\wh\E{2}=\bA\bB\bC^2\X{3}{1}$, being described as the $A_1$ singularity.
One can find a junction with zero asymptotic charges and self-intersection
$-4$ which thus generates the $u(1)$ factor although this junction is not
associated with coalescing branes.

The $\wh E_2$ curve has the singular fibers $A_1+3A_0+A_6$ in total, and
is not completely massless in the sense that the $u(1)$ factor is carried
by the non-collapsible junction. We note that
there are two possible choices of $k$ to make $\Delta$ factorized more:
\beqa
&&k=-5, \quad (H_1+A_0+A_0+A_6) \CR
&&k=51, \quad (A_1+H_0+A_0+A_6).
\eeqa
In each case, however, the singularity type remains unchanged.

\vskip3mm\noindent
$\bullet$ Massless $\wh E_1$:

Setting $a_1=0$ and $a_4=-L^4a_2^2/48$ we obtain
\beq
 y^2=x^3+\left(-{3 \over L^4}u^4 +a_2u^2
-{L^4 \over 48}a_2^2 \right) x   
+\left({2\over L^6}u^6 -{a_2 \over L^2} u^4
+{5 \over 48}L^2a_2^2u^2+{1 \over 864} L^6a_2^3 \right)
\eeq
with the discriminant
\beq
\Delta =-3 L^4a_2^4u^2 \left( 3u^2-a_2L^4 \right)/256.
\eeq
Since $\wh\E{1}=\bB\bC^2\X{3}{1}$ we have two coalescing $\bC$-branes at $u=0$,
realizing the $A_1$ singularity, and hence $E_1=A_1$.

\vskip3mm\noindent
$\bullet$ $\wh{\wt E}_1$ and $\wh{\wt E}_0$:

These cases do not admit further degeneration.

\section{Five-dimensional $E_N$ theories on $S^1$}

\renewcommand{\theequation}{4.\arabic{equation}}\setcounter{equation}{0}

In this section we consider 5D $\cN =1$ supersymmetric $SU(2)$ gauge theory
with $N_f$ quark hypermultiplets. The vector multiplet consists of a vector
field, a real scalar and a spinor, and the hypermultiplet contains four real
scalars and a spinor. $N_f$ quark hypermultiplets are in the doublet of 
$SU(2)$. Note that in 5D the bare gauge coupling $1/g_0^2$ has mass dimension 
one. When the bare quark masses vanish the global flavor symmetry
is $SO(2N_f)$. There also exists a global $U(1)_I$ symmetry generated by 
the current $j=* \Tr (F \wedge F)$ which is conserved in 5D.
Hence the massless microscopic theory possesses the global symmetry
$SO(2N_f)\times U(1)_I$.

Seiberg found that for $N_f \leq 8$ non-trivial interacting superconformal 
field theories appear in the limit $g_0 \rightarrow \infty$ \cite{Sei}. 
At these strongly-coupled
fixed points, surprisingly, the global symmetry is enhanced to $E_{N_f+1}$.
Moreover there exist two theories with global $E_1$ and $\wt E_1$ symmetry,
respectively, for $N_f=0$. The $\wt E_1$ theory further flows down to the
$\wt E_0$ theory with no global symmetry. This class of 5D
theories with exceptional global symmetry is shown to appear when M-theory
is compactified on a Calabi-Yau threefold with a vanishing four-cycle
realized by del Pezzo surfaces \cite{MS,DKV}. The correspondence between
global symmetries and del Pezzo surfaces is precisely the one presented in
Table \ref{tbl2}.

The brane web construction of various 5D theories has been
considered in the literature \cite{AH}-\cite{LV} though the $SU(2)$ theories
with exceptional symmetries were not obtained in the M-theory 5-brane
approach. In a recent interesting paper, however, it is found that
introducing 7-branes in the $(p,q)$ 5-brane web makes it possible to 
realize exceptional global symmetries on the web \cite{DHIK}.

As discussed in \cite{Nek,LN,GMS,Kol,MNW,BISTY,AHK} the Coulomb branch of 5D 
theories on $S^1$
is described in terms of complex curves where the dependence on the bare 
quark masses enters through trigonometric functions. In particular,
the curves for the $E_N$ theories on $S^1$ have been derived explicitly 
in \cite{MNW}. Their derivation of the curves is motivated by
the analysis of the local model of a shrinking del Pezzo four-cycle in a 
Calabi-Yau manifold \cite{LMW}. 

It turns out that our $\wh E_N$ curves to describe the affine 7-brane 
backgrounds are equivalent to those obtained in \cite{MNW} where $u$ is
a moduli parameter of the Coulomb branch. It is 
straightforward to compare the curves in the massless limit. 
First of all, for massless $\wh E_8$,
applying the transformation (\ref{sl2tr}) with
$c=(\frac{L}{\sqrt{3}},0,0,-\frac{18 \sqrt{3}}{b_1 L^5})$ one obtains
\beq
y^2=x^3-\frac{u^4}{3}x-2 u^5+\frac{2 u^6}{27}
\eeq
which is rewritten as
\beq
\wh E_8: \hskip10mm y^2=x^3+u^2 x^2-2 u^5
\label{lessE8}
\eeq
by a shift in $x$.

Similarly, the massless curves for $\wh E_N$ $(7 \geq N \geq 2)$ are
transformed into
\beqa
&& \wh E_7: \hskip10mm   y^2=x^3+u^2 x^2+2 u^3 x,   \CR
&& \wh E_6: \hskip10mm   y^2=x^3+u^2 x^2-2 i u^3 x-u^4,   \CR
&& \wh E_5: \hskip10mm   y^2=x^3+(u^2-4 u)x^2+4 u^2 x,   \CR
&& \wh E_4: \hskip10mm   y^2=x^3+(u^2-6iu+11)x^2+(40-40iu)x+(48-64 iu), \CR
&& \wh E_3: \hskip10mm   y^2=x^3+(u^2+10 u-23) x^2+128(1-u)x,   \CR
&& \wh E_2: \hskip10mm   y^2=x^3+(u^2+18 i u+47) x^2+512(i u-1) x-65536.
\label{lessEn}
\eeqa
Here the transformation (\ref{sl2tr}) has been utilized with 
$c=(\frac{L}{\sqrt{3}},0,0,c_4)$ where
\beq
c_4=\frac{6 \sqrt{3}}{L^3 a_1}, \quad
-\frac{6 i \sqrt{3}}{L^3 a_1}, \quad
\frac{8 \sqrt{3}}{L^3 a_1}, \quad
\frac{i T}{\sqrt{3}}
\eeq
for $N=7,6,5,4$, respectively, and
$c=(\frac{L}{\sqrt{3}},\frac{L}{\sqrt{3}},0,-\frac{4T}{\sqrt{3}})$
for $N=3$. For $N=2$ the curve has been obtained from (\ref{E2withk})
by a specialization $k=5$, a transformation (\ref{sl2tr}) with
$c=(\frac{1}{\sqrt{3}},\frac{5i}{\sqrt{3}},0,-\frac{2i}{\sqrt{3}})$
and a shift in $x$.

The curves (\ref{lessE8}) and (\ref{lessEn}) are in precise agreement with
the massless limit of the curves in \cite{MNW}.
For $\wh E_1$ and $\wh{\wt E}_1$ we take the massive curves in the form of
(\ref{massE1at}) and (\ref{masstE1at}), respectively. Setting 
$p=e^{i\lambda}$ we see that they  agree with the corresponding
curves in \cite{MNW} with $\lambda$ being the $U(1)$ mass parameter.
For massive $\wh E_{N\geq 2}$ we have not yet make a detailed comparison,
though we are sure that they will certainly agree according to the general
singularity theory.

In \cite{MNW} the $\wh E_{N}$ curves for $N\leq 7$ are derived from the massive
$\wh E_{8}$ curve by decoupling a large mass along with the scaling. The
renormalization group flows 
among the 5D $E_N$ theories follow from this analysis.
Our derivation of the generic $\wh E_{N}$ curves is simpler since what one
had to do is to generate the successive coalescence flows of the $A$-type
singularities at $u=\infty$. In order to discuss the compactification of
non-critical $E$-strings, however, expressing deformation parameters 
$a_i, b_i$ in terms of the Wilson lines is useful. 
This is carried out in \cite{MNW} by 
investigating the instanton expansion of the prepotential. We suspect that
this can also be performed in a representation theoretic way as was done
for the case of finite $ADE$ symmetries on 7-branes \cite{NTY}.

\subsection{Compactification to four dimensions}

\renewcommand{\arraystretch}{1.4}
\begin{table}
\begin{center}
\begin{tabular}{||c|c|c|c|c||} \hline
degree  & $\wh E_8$ & $\wh E_7$ & $\wh E_6$ & 
$\wh E_{N\leq 5}, \wh{\wt E}_1, \wh{\wt E}_0$ \\ \hline\hline
$q_y$   & 15    &  9    & 6     & 3         \\
$q_x$   & 10    &  6    & 4     & 2         \\
$q_u$   & 6     &  4    & 3     & 2         \\ \hline
\end{tabular}
\end{center}
\caption{Degree of variables}
\label{tbl3}
\end{table}
\renewcommand{\arraystretch}{1}
We now wish to discuss the compactification of 5D $E_N$ theories to four 
dimensions. For this purpose the degree $q_i$ ($=$ mass dimension) is assigned 
to $y,x,u$ in the curves, see Table \ref{tbl3}.
Then the scale parameters $L$ and $T^{-1}$ have mass dimension one.
We are thus led to identify $L=1/R_5$ where $R_5$ is the radius of $S^1$
along which the fifth dimension is curled up. This $R_5$-dependence of the
$u^4$ and $u^6$ terms in the curve (\ref{massE8}) agrees with the one fixed 
in \cite{GMS}. The compactification limit $R_5 \rightarrow 0$ can be taken 
directly in the massless $\wh E_{8,7,6}$ curves. We obtain 
\beqa
&& \wh E_8: \hskip10mm y^2=x^3+b_1u^5, \CR
&& \wh E_7: \hskip10mm y^2=x^3+a_1u^3x, \CR
&& \wh E_6: \hskip10mm y^2=x^3+{\alpha^2 \over 12}u^4,
\eeqa
where we have set $a_1=\alpha /L$ for $\wh E_6$ with $\alpha$ 
being kept fixed when 
$L \rightarrow \infty$. These are in the form of the $E_n$ singularities 
which describe the $\cN =2$ $E_n$ fixed points in four 
dimensions \cite{MN,NTY}. This indicates that 5D $E_{8,7,6}$ theories on
critical compactify to 4D critical $E_{8,7,6}$
theories without adjusting any relevant parameters. In view of the 7-brane
configurations the compactification limit is taken by decoupling the 
brane $\X{3}{1}$ from the $\wh \E{N}$ branes, leaving the $\E{N}$ branes
as the background to describe 4D theory.

The situation changes when we consider 5D $E_{N\leq 5}$ theories.
Taking the limit $R_5 \rightarrow 0$ of the massless curves
does not work in finding 4D theories. 
Recall that $E_N$ theories have the $N_f=N-1$ flavors. Thus,
upon compactification, 5D $E_{N\leq 5}$ theories are expected to reduce
to 4D $\cN =2$ $SU(2)$ QCD with $N_f$ flavors whose Coulomb branch is
described in terms of the Seiberg-Witten (SW) curve \cite{SW}. To see this
explicitly, we decouple $\bC$ and $\X{3}{1}$-branes simultaneously from
the branes $\wh \E{N_f+1}=\bA^{N_f}\bB\bC^2\X{3}{1}$ as $R_5 \rightarrow 0$,
leaving the branes ${\bf D_{N_f}}=\bA^{N_f}\bB\bC$ for 4D theory. 
This implies that one has to take the scaling limit
by turning on suitable mass parameters in the theory. In the following
we show how to obtain the massless SW curves for 4D theories. For this,
we recall that, in the massless case, the relevant singularity structures
on the $u$-plane are given by $D_4$ for $N_f=4$, $A_3+A_0$ for $N_f=3$,
$A_1+A_1$ for $N_f=2$, $A_0+A_0+A_0$ for $N_f=1$ and $A_0+A_0$ for $N_f=0$.
These are realized by taking the scaling limit in the massive curves for
5D theories as follows:

\vskip3mm\noindent
$\bullet$ $\wh E_5$ curve:

This case is rather simple. Setting $a_i=0$ $(i=1,3,4)$, $b_i=0$ $(i=4,5,6)$ 
and letting $L \rightarrow \infty$ we obtain
\beq
y^2=x^3+a_2u^2x.
\eeq
This is the $D_4$ singularity for the massless $N_f=4$ theory in 4D. 
The original $N_f=4$ SW curve is recovered via a change of variables as 
shown in \cite{NTY}.

\vskip3mm\noindent
$\bullet$ $\wh E_4$ curve:

In the massive curve, let us first take
\beq
a_4=-{3\over T^4}, \hskip10mm a_3=-{2 \over LT^3}\left(L^3Ta_1-18\right),
\hskip10mm b_6={2 \over T^6}, \hskip10mm b_5=-{a_3 \over T^2},
\eeq
then we are left with $a_1, a_2$. Note that $a_2$ is a dimensionless
parameter. Inspecting the curve it is observed that the desired limit is
obtained if we set
\beq
a_1={\alpha \over L^2}, \hskip10mm a_2=-{\alpha^2 \over 12}+{\beta \over TL}.
\label{scpara}
\eeq
Now, letting $L \rightarrow \infty$ yields the curve with the $A_3+A_0$
singularity. Here the scale parameter $T$ survives the compactification limit,
and hence we identify it with the QCD dynamical scale, 
$T^{-1} \propto \Lambda_3$. If we now put 
$\alpha =2, \beta =6, T=-i\sqrt{3} \Lambda_3$ and make a shift in $x$,
the massless $N_f=3$ SW curve follows
\beq
y^2=x^2(x-u)-{\Lambda_3^2 \over 64}(x-u)^2.
\eeq

\vskip3mm\noindent
$\bullet$ $\wh E_3$ curve:

We take $a_4=-3/T^4, b_6=2/T^6, a_3=-1/T^2$ and set
\beq
a_1={\alpha \over L^2}, 
\hskip10mm a_2=-{\alpha^2 \over 12}+{\beta \over (TL)^2}.
\eeq
Letting $L \rightarrow \infty$ yields the curve with the $A_1+A_1$
singularity. the scale parameter $T$ is again converted into the QCD scale
$\Lambda_2$. If we now put $\alpha =2, \beta =-6, T=2\sqrt{3}/\Lambda_2$
and make a shift in $x$ as well as in $u$, 
the massless $N_f=2$ SW curve follows
\beq
y^2=x^2(x-u)-{\Lambda_2^4 \over 64}(x-u).
\eeq

\vskip3mm\noindent
$\bullet$ $\wh E_2$ curve:

Setting $a_3=a_4=0$ and
\beq
a_1={\alpha \over L^2}, 
\hskip10mm a_2=-{\alpha^2 \over 12}+{\beta \over (TL)^3},
\eeq
we let $L \rightarrow \infty$. The resulting curve exhibits the singularity
$A_0+A_0+A_0$. Putting $\alpha =2, \beta =i3\sqrt{3}/4$ and $T=1/\Lambda_1$
we obtain the massless $N_f=1$ SW curve
\beq
y^2=x^2(x-u)-{\Lambda_1^6 \over 64}.
\eeq

\vskip3mm\noindent
$\bullet$ $\wh E_1$ curve:

Setting $a_4=1/(4T^4)$ and
\beq
a_1={\alpha \over L^2}, 
\hskip10mm a_2=-{\alpha^2 \over 12}+{\beta \over (TL)^4},
\label{pure}
\eeq
we let $L \rightarrow \infty$. The resulting curve exhibits the singularity
$A_0+A_0$, Putting now $\alpha =2, T=1/\Lambda_0$ ($\beta$ remains arbitrary)
yields 
\beq
y^2=x^2(x-u)+{\Lambda_0^4 \over 4}x.
\label{pureSW}
\eeq
This is the SW curve for $\cN =2$ $SU(2)$ pure Yang-Mills theory.

\noindent
$\bullet$ $\wh{\wt E}_1$ curve:

We set $a_3=0$ and (\ref{pure}), and then 
let $L \rightarrow \infty$, yielding the curve with the $A_0+A_0$
singularity. Putting now $\alpha =2, \beta =-9/2, T=1/\Lambda_0$ we 
get the SW curve (\ref{pureSW}) as in the previous case.

Thus we have seen that there exists the compactifiction limit of
5D $\cN =1$ $E_{N\leq 5}$ theories on $S^1$ down to 4D $\cN =2$ $SU(2)$
QCD with $N_f=N-1$ flavors. For $N\leq 4$ the compactifiction limit is
taken as the scaling limit as prescribed in (\ref{scpara}). The appearance of 
the QCD scale $\Lambda_{N-1}$ reflects the fact that the $\E{N}$ brane
configurations for $N\leq 4$ are not collapsible. It will also be possible 
to derive the massive $SU(2)$ SW curves from the massive $\wh E_{N}$ curves,
see ref.\cite{Gan-2} for earlier related computations 
in view of 6D non-critical strings.

\section{Discussion}

Starting with a rational elliptic surface for $\wh \E{9}$ 
we have constructed the the elliptic curves corresponding to the affine 
7-branes $\wh \E{N}$ $(1 \leq N \leq 8)$ and
$\wh{\wt \bE}_{\bf N}$ $(N=0,1)$. The brane picture is quite efficient in
working out these curves explicitly. We have then shown that the curves for
the affine 7-brane backgrounds describe the Coulomb branch of the 5D
$E_N$ theories compactified on a circle. The result indicates that the idea
of the D3-brane probe is valid for the description of 5D $E_N$ theories on
$\bR^4 \times S^1$; both a probe D3-brane and background 7-branes extend over
the bulk $\bR^4$, and the dependence on $S^1$ is encoded in the ``affinizing''
7-brane $\X{3}{1}$. The BPS states arise from $(p,q)$ strings/junctions
stretching between the D3-brane and 7-branes. Thus there will exist BPS states
with non-zero magnetic gauge charge $q \not= 0$ in 5D theory on 
$\bR^4 \times S^1$. On the other hand, the BPS states of the 5D $E_N$ theory
in the bulk $\bR^5$ are represented as the 5-7 strings with charges 
$(p,q)=(2n_e, 0)$ where $n_e$ is the Cartan charge of 5D $SU(2)$ 
gauge group \cite{DHIK}. The level $k$ of the Kac-Moody algebra realized on the
affine exceptional 7-branes (\ref{canoE8}) is related to $q$ 
through $k=-q$ \cite{dW,DHIZ-3}. Therefore only the finite part of $\wh E_N$
is relevant to the enumeration of BPS states in the bulk 5D theory.
As just mentioned above, however, 
the 5D theory on $S^1$ enables to carry BPS states
with $q \not= 0$, and hence the affine property may play a manifest role.
In fact, interesting affine $\wh E_N$ structures have already been revealed
in the context of non-critical $E$-strings \cite{KMV,MNW,MNVW}. It will be
worth pursuing further the idea in the framework of the 7-brane setup.

Mathematically speaking, what we have investigated in this paper is the 
problem of understanding the moduli of rational elliptic surfaces 
(\ref{cubic}), which is a space of polynomials
\beq
f(z)=\sum_{i=0}^4 a_iz^i, \hskip10mm g(z)=\sum_{i=0}^6 b_iz^i
\eeq
divided by (\ref{sl2tr}).
Since the moduli space consists of various branches with different
singularity structures, it is important to consider the
interrelation (degeneration-coalescence) among the strata \cite{Mir,MP}.
The brane picture provides a natural description of
such degeneration-coalescence structures.

The whole construction presented in this paper may be 
summarized in the degeneration-coalescence diagram. The degeneration
diagram turns out to be
\beq
\begin{array}{ccccccccccccccccccccc}
&&&&&&&&&&&&&&\wh{\wt{E}}_1&\rr&\wh{\wt{E}}_0 \\
&&&&&&&&&&&&&\ne \\
\wh{E}_8&\rr&\wh{E}_7&\rr&\wh{E}_6&\rr&\wh{E}_5&
\rr&\wh{E}_4&\rr&\wh{E}_3&\rr&\wh{E}_2&\rr&\wh{E}_1 \\
\downarrow&&\downarrow&&\downarrow&&\downarrow&&\downarrow&&
\downarrow&&\downarrow&&\downarrow \\
E_8&\rr&E_7&\rr&E_6&\rr& (E_5&\rr&E_4&\rr&E_3&\rr&E_2&\rr&E_1) \\
&&&&&&\downarrow&&\downarrow&&\downarrow&&\downarrow&&\downarrow \\
&&&&&&D_4&\rr&(D_3&\rr&D_2&\rr&D_1&\rr&D_0) \\
&&&&&&&&\downarrow&&\downarrow&&\downarrow \\
&&&&&&&&H_2&\rr&H_1&\rr&H_0 
\label{dege}
\end{array}
\eeq
In this diagram, where each arrow stands for the flow generated
by sending a single 7-brane to infinity,  the $\wh E$-sequence represents
the 5D $\cN =1$ $E_N$ theories. Then the down arrows from the $\wh E$ to
the $E$-sequence correspond to the compactification to four dimensions.
The flavor number $N_f=N-1$ is preserved along down arrows for each $N$.
The $E$-sequence as well as the $D$- and $H$- sequences are 4D $\cN =2$ 
theories with various global symmetries where the theories put in the brackets
are not classified as non-trivial fixed points. 

Corresponding to the degeneration flows (\ref{dege}), we have the coalescence 
flows among the singularity structures on the opposite side of the base 
$\bP^1$. These flows are shown in the following coalescence diagram:
\beq
\begin{array}{ccccccccccccccccccccc}
&&&&&&&&&&&&&&A_7&\rr&A_8 \\
&&&&&&&&&&&&&\ne \\
A_0&\rr&A_1&\rr&A_2&\rr&A_3&\rr&A_4&\rr&A_5&\rr&A_6&\rr&A_7  \\
\downarrow&&\downarrow&&\downarrow&&\downarrow&&\downarrow&&
\downarrow&&\downarrow&&\downarrow \\
H_0&\rr&H_1&\rr&H_2&\rr&H_4&\rr&H_5&\rr&H_6&\rr&H_7&\rr&H_8 \\
&&&&&&\downarrow&&\downarrow&&\downarrow&&\downarrow&&\downarrow \\
&&&&&&D_4&\rr&D_5&\rr&D_6&\rr&D_7&\rr&D_8 \\
&&&&&&&&\downarrow&&\downarrow&&\downarrow \\
&&&&&&&&E_6&\rr&E_7&\rr&E_8 
\label{coal}
\end{array}
\eeq
Each entry of this diagram represents the singularity type.

In order to analyze the BPS spectrum of 5D $E_N$ theories on $\bR^4 \times S^1$
one turns to the study of the junction lattice on the 7-branes. For this
purpose it is desired to have a further understanding of how the junction 
lattice is described in terms of geometry of rational elliptic surfaces.
This amounts to analyzing the sections of the massive $\wh E_n$ curves based
on the brane picture and its relation to the Mordell-Weil lattice \cite{FYY}.
The analysis will also become interesting from the standpoint of the
F-theory/heterotic duality in eight dimensions.

\vskip4mm\noindent
{\bf Acknowledgements}

\vskip2mm
We would like to thank M. Fukae for discussions and B. Zwiebach for useful
comments on the manuscript.
The work of SKY was supported in part by Grant-in-Aid for Scientific Research 
on Priority Area 707 ``Supersymmetry and Unified Theory of Elementary 
Particles'', Japan Ministry of Education, Science and Culture.

\vskip6mm\noindent
{\bf Note added}

\vskip2mm
Just after this paper was submitted, an interesting paper by A. Sen 
and B. Zwiebach appeared \cite{SZ} in which they have also constructed the 
elliptic curves for the affine 7-branes. Their construction substantially
overlaps with ours in section 3.

\newpage


\end{document}